\documentclass{ws-ijmpc}

\begin{document}

\markboth{P\'eter Lajk\'o}
{Renormalization-group investigation of the $S=1$  random antiferromagnetic Heisenberg chain }

\catchline{}{}{}{}{}
%%%%%%%%%%%%%%%%%%%%%%%%%%%%%%%%%%%%%%%%%%%%%%%%%%%%%%%%%%%%%%%%%%%%%

\title{Renormalization-group investigation of the $S=1$  random antiferromagnetic Heisenberg chain
}

\author{P\'eter Lajk\'o}

\address{Department of Physics, Kuwait University\\
Kuwait City, Safat 13060, P.O. Box 5969,
Kuwait\\ lajko@physics.kuniv.edu.kw}
   
\maketitle

\begin{history}
\received{Day Month Year}
\revised{Day Month Year}
\end{history}

\begin{abstract}
We introduce variants of the Ma-Dasgupta renormalization-group (RG) approach for random quantum spin chains,
in which the energy-scale is reduced by decimation built on either perturbative
or non-perturbative principles. In one  non-perturbative version
of the method, we require the exact invariance of the lowest
gaps, while in a second class of perturbative Ma-Dasgupta techniques,
different decimation rules are utilized. 
For the $S=1$ random antiferromagnetic Heisenberg chain, both type of methods provide the same type of disorder dependent
phase diagram, which is in agreement with density-matrix renormalization-group
(DMRG) calculations and previous studies. 
\end{abstract}

\ccode{PACS Nos.:  05.50.+q, 64.60.Ak, 68.35.Rh}

\section{Introduction}
In recent years intensive research work has explored some curious properties
of antiferromagnetic Heisenberg chains.\cite {Haldane,Hallberg,Hallberga,Ma,Maa,Fisher} The features of the pure systems are
well-known after Haldane's seminal work:\cite{Haldane} half-integer $S$
and integer $S$ chain show distinct behavior;  gapless spectra, quasi-long-range
order and  gapped spectra, hidden topological order characterize them,
respectively. In  $S=1/2$ chain, Ma and Dasgupta introduced a RG
technique,\cite{Ma,Fisher} with which  it was possible to point out that any
small amount of disorder triggers the 
random-singlet phase (RSP) where spins in arbitrarily large distance
are coupled in singlets. In  $S=1$ chain,\cite{Monthus,Monthusa,Yang,Yanga,Hida,Con,Todo,Todoa,Todob}  only sufficiently large disorder is able to change the pure
system properties. 

The  $S=1$ random antiferromagnetic Heisenberg chain (RAHC) was studied 
numerically with density-matrix renormalization-group (DMRG)
\cite{Hida,DMRG1,DMRG1a,DMRG1b,DMRG2,DMRG2a} and quantum Monte Carlo (QMC)
 technique,\cite{Todo,Todoa,Todob,Berg}
a considerable part of the findings was contributed by extended
 versions\cite{Monthus,Monthusa,Yang,Yanga} of the  Ma-Dasgupta RG method that    is based on a perturbative decimation step, which fails for
 higher-S RAHC for weak disorder, but
 asymptotically correct for strong disorder, therefore, it is known as
 strong-disorder RG (SDRG). The SDRG method has been applied in chains,\cite{Lin,Lina}
 ladders,\cite{Melin}
  in two-, and three-dimensional systems,\cite{Lin2} for a review by
 Igl\'oi and Monthus, see Ref. \refcite{Ferenc}.

In this paper, a non-perturbative RG technique is presented, which is inspired by the
 SDRG method\cite{Ma,Maa} and in which the lowest gaps play the central role,
 therefore it is termed as gap RG (GRG). 
The original SDRG is investigated: a simple argument is presented and justified, by using
 local and global aspects of disorder, to explain the 
 success of the SDRG, which holds true in higher-$S$ chains.\cite{Lajko}   

 The structure of the paper is the following: in Section 2, the
 model and the failure of the SDRG decimation are described. The results of the
 GRG and the disorder-induced phases are described in Section 3. Different
 SDRG methods, the interpretation and justification of their correctness,  are detailed in Section 4.
 The results are summarized and compared to earlier findings, in Section 5.

\section{The random antiferromagnetic Heisenberg chain and the strong-disorder renormalization}

We investigate the $S=1$ RAHC with Hamiltonian 

\begin{equation}
H = \sum_{i=1}^{L-1} J_{i} \vec{S}_i \cdot \vec{S}_{i+1}.
\end{equation}
The $J_i>0$ couplings follow a random distribution,  the power-law distribution 
 
\begin{equation}
p_\delta (J) = \delta^{-1} J^{-1+1/\delta}  \ \ \ \ \ {\rm for} \ 0 \leq
J \leq 1 \ \  \label{distribut}
\end{equation}
where the $\delta$ parameter tunes the strength of disorder as $\delta^2= \rm
var[\ln \it J]$;
$\delta=0$, $\delta=1.0$, and $\delta\to \infty$ denote pure, uniform, and
infinite randomness, respectively.
This system has a relevance in solid state physics,\cite{Fazekas} in quantum
computations,\cite{Chris} and serves as a testing ground for 
 developing novel methods and theories.\cite{Enrico,IFL} 

\begin{figure}
\centering
\includegraphics*[width=5.5cm]{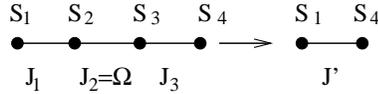}% Here is how to import EPS art
\caption{\label{fig:epsart}The four-spin system with the largest coupling parameter $\Omega$ replaced by a two-spin system with  coupling constant given in Eq.(\ref{J_pert}), in a usual SDRG decimation, or  such $J'$ which involves an identical first gap as that  of the original four-spin subchains, in a GRG decimation.}
\end{figure}

In a given RAHC, the SDRG method  eliminates iteratively the largest coupling with its four neighboring spins and
replaces it by two spins with a new coupling, see Fig. \ref{fig:epsart}, determined by perturbation theory
for general $S$ as

%\section{Gaps}

\begin{equation}
J'=\frac{2}{3} \frac{S(S+1)J_{1} J_{3}}{\Omega}\;.
\label{J_pert}
\end{equation}
The method works in $S=1/2$
             RAHC where the new coupling $J'=J_{1}J_{3}/2\Omega$
             is  smaller than any of the
             eliminated ones. For higher-$S$ RAHC the
  $(2/3)S(S+1)>1$, the new
             coupling can be larger than
  the decimated one: the method fails for weak disorder. 
In order to circumvent this problem, novel RG schemes were
initiated\cite{Monthus,Monthusa,Yang,Yanga}  that use a set of decimation steps and deal with a
mixture chain of $S=1/2$ and $S=1$ spins.
 These RG schemes predict a transition
from the pure system behavior to the RSP, which was tested with  DMRG\cite{Hida} and QMC.\cite{Berg,Todoa}   
Saguia et al.\cite{Con}  use a different strategy to tackle the problem, they keep the
original SDRG decimation, if it works correctly, and replace it
with a two-step perturbative decimation, if it fails. The method yields three
phases: gapped/gapless Haldane phase (or nonsingular/singular Griffiths phases), and RSP.

\section{Gap renormalization-group method}

We briefly describe a non-perturbative RG method that has a similar iterative strategy as
the SDRG: it eliminates the largest couplings,
but the decimation rule is different, the new coupling is determined with the condition that the first gap is invariant, see
Fig. \ref{fig:epsart}. 
The system properties can be drawn from the first gap
distribution and by analyzing the RG flows. 
The GRG works in higher-$S$ RAHC and in
 $S=1/2$ chain in its original form.\cite{Lajko}  Preliminary results of GRG
were published in Ref.\refcite{Lajko2}. In the following, more detailed
investigations  are presented.

\begin{figure}
\centering
\includegraphics*[width=6.6cm]{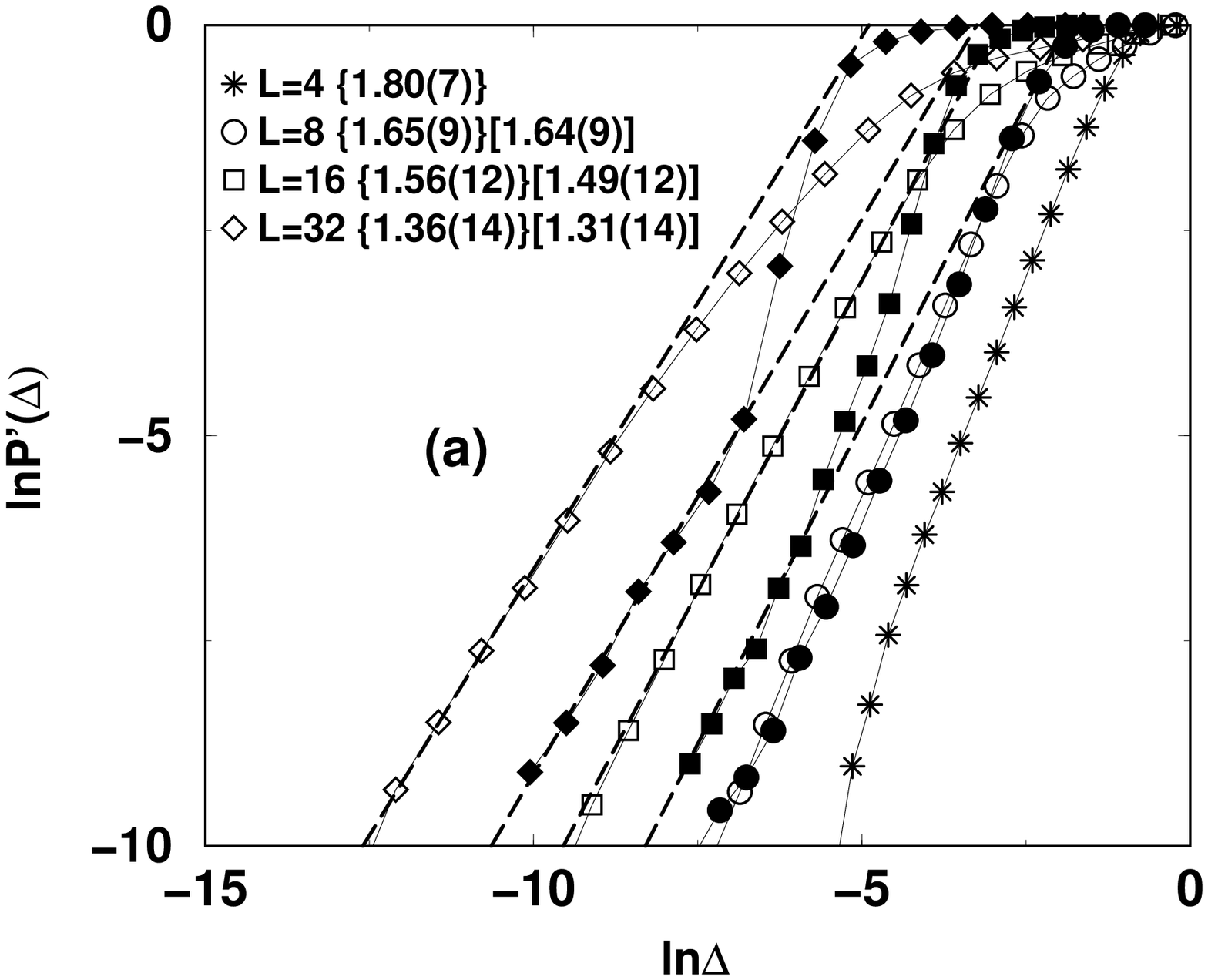}%
\hspace{0.1cm}%
\includegraphics*[width=5.7cm]{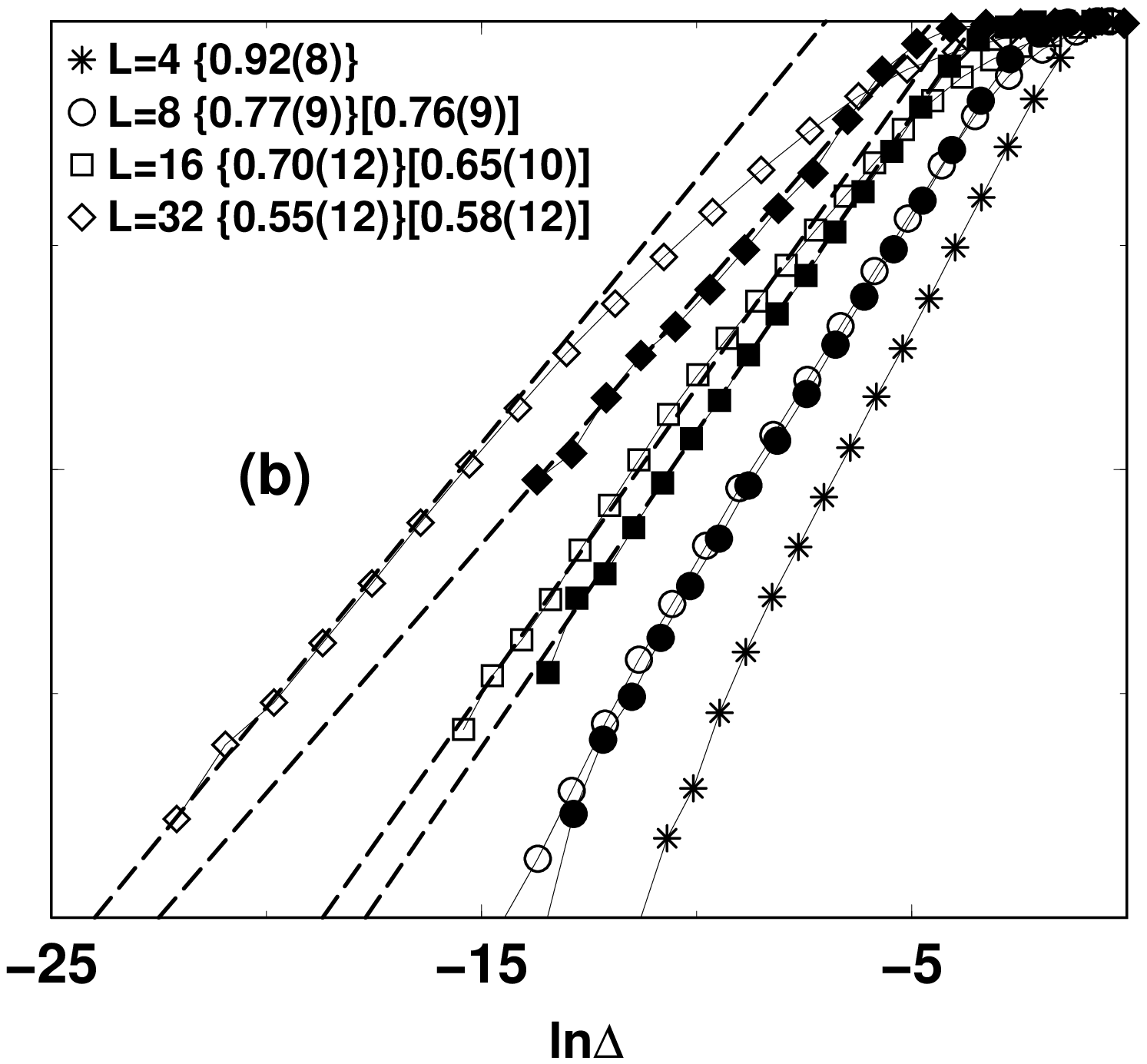}% Here is how to import EPS art
\caption{\label{fig:epsart5} Integrated probability distribution of the first gaps at
  $\delta=0.5$ (a) and  at $\delta=1.0$ (b) disorder. In terms of the slopes
  of the low-gap region, the agreement is obvious, although the GRG
  data (empty symbols) are  shifted toward the low-gap region. DMRG data (filled symbols) are
  available up to $L=32$,  the GRG data are plotted also up to this length. The
  estimated slopes are indicated in the legends (i.e., \{DMRG slopes\} and
  [GRG slopes]) and plotted with dashed lines. }
\end{figure}

\subsection{The disorder-induced phases and finite-size corrections}
We have investigated the $S=1$ RAHC by
    using the GRG technique and DMRG\cite{DMRG1,DMRG1a,DMRG1b,DMRG2,DMRG2a} method
    in parallel. 
We eliminated iteratively  $L-4$ spins from the original $L$  by using the GRG
method, the new
couplings were always smaller than the eliminated ones, and the final four-spin system
is diagonalized by the L\'anczos method. On the other hand, genuine first gaps
are calculated by using DMRG method. 
\begin{figure}
\centering
\includegraphics*[width=7.5cm]{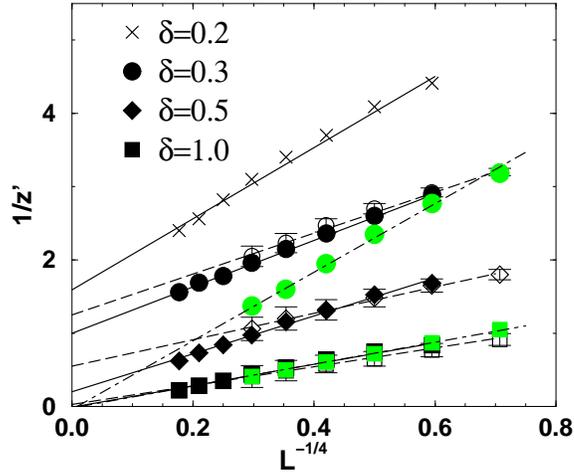}% Here is how to import EPS art
\caption{\label{fig:epsart7} The $1/z'$ exponent. Empty symbols (dashed lines) denote the GRG
  data (fitted straight lines). Full symbols (solid lines)
  denote the SDRG data. Opaque symbols (dot-dashed lines) denote the XX chain data. Error bars, if not indicated, are smaller than the symbols.}
\end{figure}
The basis number in DMRG calculations was so large (maximum $180$) that it did
not influence the probability distribution. 

The comparison was made for different strength of
disorder, always a very good agreement was found between the original
 gap integrated probability distribution and the GRG data in the sense
 that the slopes of the low-gap tail from GRG and DMRG are the same, see in Fig. \ref{fig:epsart5}.     In fact, this is the first
time that  RG results of $S=1$ RAHC are  faced to genuine probability
distributions.

The dynamical exponent $z$ can be determined via the scaling of the first
gap probability densities (and integrated probability densities) in the low-gap region:
\begin{equation}
P(\ln\Delta) \sim \Delta^{1/z'},
\label{omegae}
\end{equation}
where $z'$ is identical to the dynamical exponent $z$, if $z'>1$, otherwise
 $z=1$.    
From the low-gap  tail of the integrated probability distributions the slopes
 ($1/z'$) in the log-log plot
are extracted for different sizes and $\delta$ parameter,
Fig. \ref{fig:epsart7}.  In this parameter region, the
following relation is conjectured and utilized in the figure  
\begin{equation}
 \frac{1}{z'(\delta,L)} \approx
 \frac{1}{z'(\delta,\infty)}+A(\delta)L^{-1/4}.      \label{distribut31}
\end{equation}
$1/z'(\delta,\infty)$ helps to identify the phase diagram of the $S=1$ RAHC. At
very weak disorder, up to $\delta_1=0.4(15)$,  $z'(\delta,\infty)$ is smaller
than $1$ indicating the nonsingular
Griffiths phase.\cite{Lajko2} At intermediate disorder $0.4<\delta<1.0$, $z'(\delta,\infty)$ has still finite
value, singular Griffiths phase, while above  $\delta_c= 1.00 (15)$ we have found
$z'=\infty$ in the large-$L$ limit indicating the appearance of RSP.

\begin{figure}
\centering
\includegraphics*[width=5.cm]{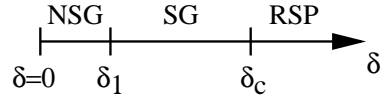}% Here is how to import EPS art
\caption{\label{fig:epsart61} The phase diagram. $\delta=0$
  denotes the pure system, $\delta_1$ the phase boundary between the
  nonsingular and singular Griffiths phases, $\delta_c$ the phase boundary
  between Griffiths phase and RSP. }    
\end{figure}

We have performed SDRG also on XX chains and
analyzed the gap distributions in the same size region.
The XX chain is chosen because its decimation rule,\cite{Fisher2} with 1 prefactor in Eq.(\ref{J_pert}), is the closest to the decimation
rule of $S=1$ RAHC, with prefactor $4/3$.  The slopes,
extracted from the XX chain at $\delta=0.3$ and $\delta=1.0$ and plotted in
Fig. \ref{fig:epsart7}, 
scale with effective exponent $-1/4$ and predict disappearing $1/z'$, as it is
expected,\cite{Fisher}  and thus justify the
assumption that these system sizes provide physically conclusive results in
the large-$L$ limit. 

\begin{figure}
\centering
\includegraphics*[width=7.cm]{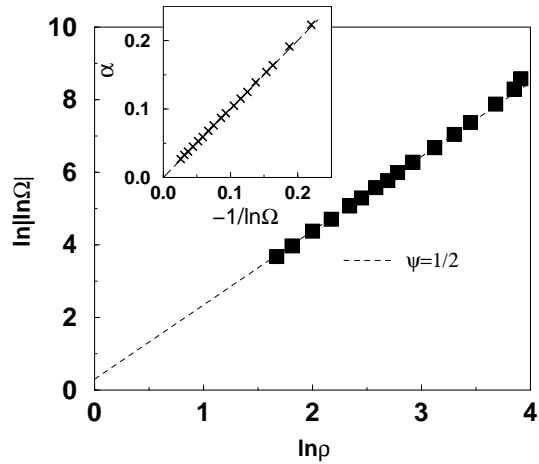}% Here is how to import EPS art
\caption{\label{fig:epsart2} Analysis of the GRG flow at $\delta=1.0$. The $\psi$ exponent from the density
  of active spins is identified from the slope as $1/2$. Inset: The $\alpha$
  parameter can be identified as  $\alpha=-1/\ln\Omega$.}
\end{figure}

\subsection{ Gap renormalization flow in the RSP}
In RSP during the RG treatment of a certain random chain, the
ratio of the remaining spins that are not decimated out follows the scaling
 law\cite{Fisher}
\begin{equation}
\rho= \frac{1}{L} =  \frac{1}{ |\ln\Omega|^{1/\psi}}.      \label{distribut3}
\end{equation}
This relation describes the connection between the characteristic
length scale of the system $L$ and the energy scale $\Omega$ via the $\psi=1/2$
exponent, a characteristic relation of RSP. The GRG method is applied on very
large system sizes, $20$ million
spins, results are plotted in
Fig.\ref{fig:epsart2}.  The slope of the double logarithmic plot
reveals the presence of RSP in the system at $\delta=1.0 $ disorder.

The following aim is to show
out that the scaling of the coupling distribution under GRG 
procedure follows the expected RSP behavior:\cite{Fisher} 

\begin{equation}
P_\Omega (J) = \frac{\alpha}{\Omega}  \left( \frac{\Omega}{J} \right)^{1-\alpha}     \label{distribut2}
\end{equation}
with $\alpha=-1/\ln\Omega$.
Actually, the above type of scaling is found at
$\delta=1.0$, see the inset in Fig.\ref{fig:epsart2} where $\alpha$,
determined from the slope of the coupling distribution, is depicted as a
function of $-1/\ln\Omega$.    
Both quantities, the coupling distribution and the fraction of undecimated spins
follow  the scaling that is characteristic of the RSP.

\section{Results of the SDRG}

In this section, different aspects of
SDRG method are analyzed and discussed. Firstly, proper interpretation of the
strong-disorder limit is presented that explains the reliability of SDRG
method. Secondly, the scaling
relations and the corresponding finite-size corrections of SDRG data are
discussed. Finally, an unusual representation of the SDRG flows is established
that has a very high accuracy.  

\begin{figure}
\centering
\includegraphics*[width=7.5cm]{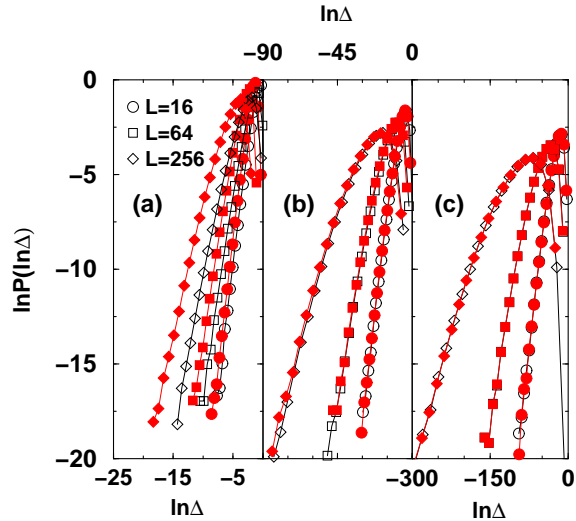}% Here is how to import EPS art
\caption{\label{fig:epsart71} (Color online) The gap distributions generated by
  the original SDRG (empty symbols) and the modified SDRG (filled symbols) for $\delta=0.3, 1.0, 3.0$ disorder
  in (a), (b), and (c), respectively. 
}
\end{figure}

\subsection{The low-gap region of the probability distributions}

One has to clearly distinguish two different kinds of appearance of disorder:
global and local appearances. On the one hand, the
global disorder describes the statistical ensemble of couplings in the chains and it can be
parameterized, in this work, with the $\delta$ parameter of the power-law
distribution. On the other hand, the local disorder or subchain disorder characterizes only one
particular four-spin subchain and it can be
parameterized, for instance, with the value $\max(J_1,J_3)/\Omega$. 

The SDRG can be correct in the proper strong-disorder
limit: if the subchain disorder is strong everywhere in a chain that is
already a sufficient condition. 
In certain chains the strong subchain or local disorder can be present generally at arbitrary location in the
chain with some small
probability that itself depends on the global disorder that describes the
statistical ensemble of the couplings.
These probably rare chains provide the smallest gaps to the gap distributions since
in these chains repeatedly very small new couplings are replaced in the
consecutive 
decimations resulting  the smallest final couplings and gaps.    
Consequently, the low-gap regime of the results can be correct, in RG sense,
at any given strength of global disorder, as they come via
the correct decimation steps from those rare chains in which
the subchain disorder is strong everywhere. This interpretation of the low-gap region is in accordance with
the well-established scenario of rare region effects.\cite{Igloi3,Vojta}    

\subsection{Comparison of the original and the modified SDRG}
In order to make this feature more transparent, we
present a systematic investigation and comparison of SDRG and modified SDRG
results in the low-gap region of the probability distributions. The original
SDRG method is straightforward, the modified SDRG applies the same decimation rule except
in those cases in which the new coupling $J'$ is larger than at least one of $J_1$
and $J_3$. In those
decimation steps $cJ'$, with $c\le 1/[(2/3)S(S+1))]=3/4$,  is used as a new
coupling ensuring always smaller new coupling than the decimated ones.  The two
schemes are  almost identical, the only difference is factor $c$ if the
perturbative decimation fails. 

This modified version of SDRG has a similar
scenario as the RG method applied in Ref. \refcite{Con}. However, the modification is
very simple here, just a rescaling of the coupling given by the original rule,
whereas a two-step degenerate perturbative decimation isis  applied in Ref.\refcite{Con}. Nevertheless, the heuristic argument used here to explain the
success of SDRG applies also for the modified version SDRG in Ref. \refcite{Con}.   
 
These two RG schemes, the original SDRG and the modified one have been applied to generate the
gap distribution for several system sizes and strength of disorder. The two
 types of gap distributions are plotted in Fig. \ref{fig:epsart71}; the slopes of the
low-gap regions are identical for fixed sizes and strengths of global disorder. The plotted results
were generated with $c=2/3$ in the modified decimation steps, but several other values provided identical slopes
in the low-gap region. 

The slope of the low-gap region is apparently
independent of those decimation steps where the original decimation rule fails in
perturbative sense.
 The modified RG steps, by generating smaller coupling than the original one, shift the corresponding low-gap distributions towards the lower
gaps, but the structure of this regime remains the same as in the
original case. A similar relation of the GRG and DMRG results is observed in
Fig. \ref{fig:epsart5}. For longer chains that involve more decimation steps, the
shift between the two gap distributions is more enhanced, see Fig. \ref{fig:epsart71}.
For sufficiently strong global disorder the modified decimation steps practically do
not make change on the structure of the gap distributions, at least for smaller chain sizes,
see Fig. \ref{fig:epsart71}(b) and (c) for $\delta=1.0$ and $\delta=3.0$
disorder, respectively.

\subsection{Disorder-induced phase diagram}
The original SDRG results were carefully analyzed and the estimates of $1/z'$
were extracted for different system sizes and 
plotted in Fig.\ref{fig:epsart7} as a function of $L{^{-1/4}}$. These
numerical values are in agreement with the results 
of GRG method. The estimates for the phase boundaries $\delta_1=0.29(6)$, $\delta_c=1.0(1)$ are in rather good agreement with other studies.  

One has to notice, the finite-size corrections are smaller in the Griffiths
phase than in the RSP, in terms of the dynamical exponent $z$, and are weakening
deeper into Griffiths phase. The SDRG results  are more reliable than
the GRG results.
  Firstly, more random chains are averaged for SDRG, at
least 80 million,  and only 2 million for the
GRG. Secondly, larger systems are analyzed for the SDRG ($L=1024$ and
only $L=128$ for GRG). 
Even larger system sizes were tested, however for those
sizes the convergence in the low-gap region was not satisfactory although
  intensive computations were
performed on a Beowulf Linux cluster in order to have accurate
gap distributions from GRG, SDRG, DMRG methods.

\begin{figure}
\centering
\includegraphics*[width=8.5cm]{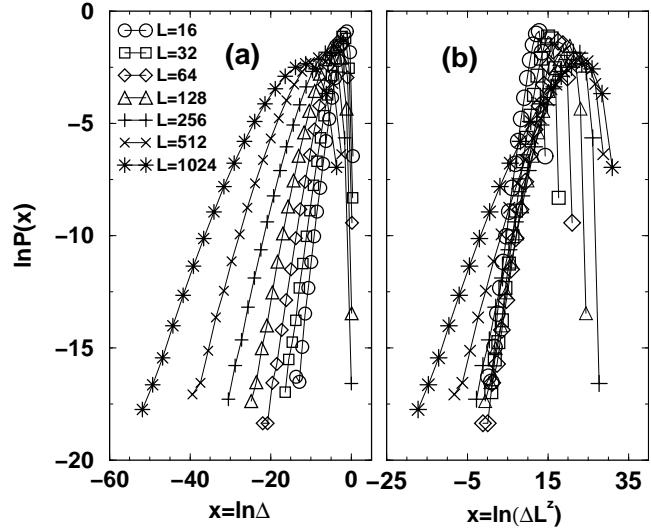}% Here is how to import EPS art
\caption{\label{fig:epsart10}(a) Gap distribution generated by the original SD\
RG
  method at $\delta=0.5$. (b) Scaling collapse assuming Griffiths phase
  with scaling relation in  Eq.(\ref{distribut4}) and  $z=5.0$.}
\end{figure}

\subsection{Scaling of the SDRG data}
In the following, the scaling of the SDRG results in the $S=1$ RAHC are tested by the expected
scaling laws\cite{Fisher} in the Griffiths phase and in the RSP that are shown out exactly, for instance, in the random transverse-field Ising spin chain, in the  $S=1/2$ RAHC or in the random XX chain. In the Griffiths phase, the
energy scale $\Omega$  is continously reducing during the
RG, and the
couplings follow the distribution   
\begin{equation}
P (J,\Omega) dJ= \frac{1}{z}  \left( \frac{J}{\Omega} \right)^{-1+1/z}  \frac{dJ}{\Omega}.   \label{distribut32}
\end{equation}
    Due  to the relation of energy scale and the length of the system (
$\Omega \sim L^{-z} $) and  by assuming that the RG process leaves the $z$
exponent unaltered, one can conclude easily Eq.(\ref{omegae}) and  the
scaling law

 \begin{equation}
P_L (\ln\Delta) = P(\ln(\Delta L^z)).   \label{distribut4}
\end{equation}
In the RSP, where $z$ is infinity, one can set formally $z=-\ln\Delta\sim L^{\psi}$, Eq.(\ref{distribut3}) transforms
into Eq.(\ref{distribut2}), and the corresponding scaling relation is
\begin{equation}
P_L (\ln\Delta) = L^{-\psi}P(\ln(\Delta)/L^{\psi}))   \label{distribut5}
\end{equation}
with $\psi=1/2$. These scaling laws were successfully utilized, for instance, in the random XX spin chains,\cite{Juhasz} in the $S=1/2$ RAHC\cite{Fisher2b} and also in higher-dimensional systems,\cite{Lin2} but were never tested in the $S=1$ RAHC due to the known reasons.

On the one hand, these scaling laws assume that the investigated sizes are
free of finite-size corrections. If this condition is not accomplished, and it is not accomplished in the $S=1$ RAHC, one can
expect the appearance of finite-size corrections in addition to the above detailed
leading behavior. On the other hand, Eq.(\ref{distribut32}) assumes the
monotoniicc decrease of the energy scale $\Omega$ that is surely not rigorously
true during the original SDRG in the $S=1$ RAHC, therefore
Eq.(\ref{distribut32}) is surely not accurately true. However,
Eq.(\ref{distribut32}) can be true in a looser sense, i.e., its consequences,
Eq.(\ref{omegae}), Eq.(\ref{distribut3}), Eq.(\ref{distribut4}) and
Eq.(\ref{distribut5}), can be true since the extraction of these relations
does not assume the rigorous accurateness of Eq.(\ref{distribut32}). The
probability distribution in the low-gap region follows the expected behavior,
Eq.(\ref{omegae}), as shown in the earlier subsection. The scaling
behavior in the Griffiths phase, Eq.(\ref{distribut4}), and in the RSP, Eq.(\ref{distribut5}), are investigated in the following paragraphs. 

The SDRG gap distributions are  plotted in  Fig.\ref{fig:epsart10} for
$\delta=0.5$ disorder. One can notice a
systematic broadening in the distributions and systematic change in the slopes of the
low-gap region with increasing systems sizes, which can be considered as a  
 result of finite-size corrections. The presence of finite-size corrections
is evident also from Fig. \ref{fig:epsart7}, where the estimated slopes are
plotted. Due to these
finite-size corrections the expected scaling collapse is not perfect, see Fig.\ref{fig:epsart10}(b). It is to be
emphasized that the values of the $z$ exponent in the scaling collapse are taken
from the large-$L$ estimates in Fig.\ref{fig:epsart7} and they are not
optimized for the scaling collapse itself. 
  In order to have even better scaling collapse  by using the same
large-$L$ limit estimates of $z$  or to see
a satisfying fit with an analytically estimated gap distributions,\cite{RYF} one should
study drastically larger systems sizes.

\begin{figure}
\centering
\includegraphics*[width=8.3cm]{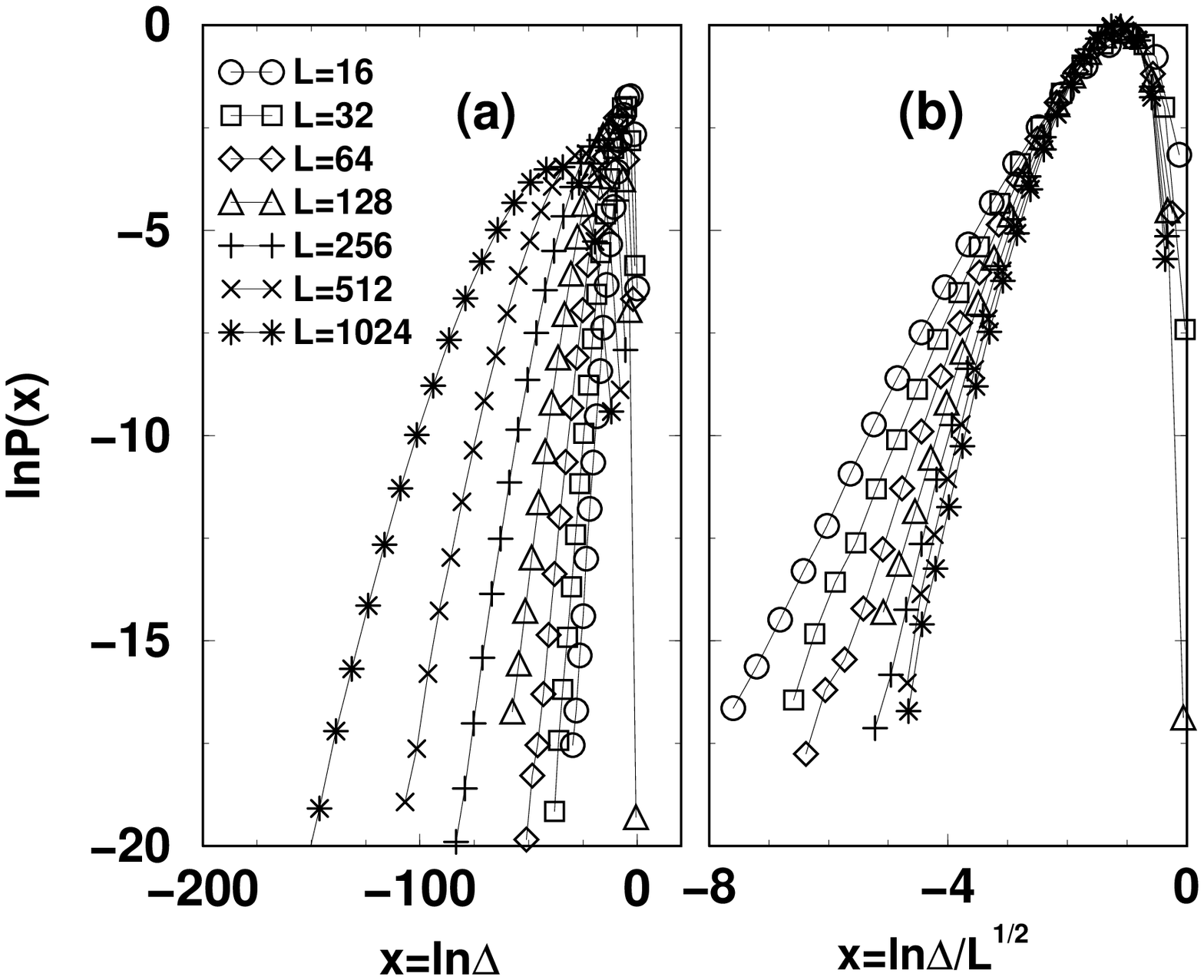}% Here is how to import EPS art
\hspace{0cm}
\includegraphics*[width=3.72cm]{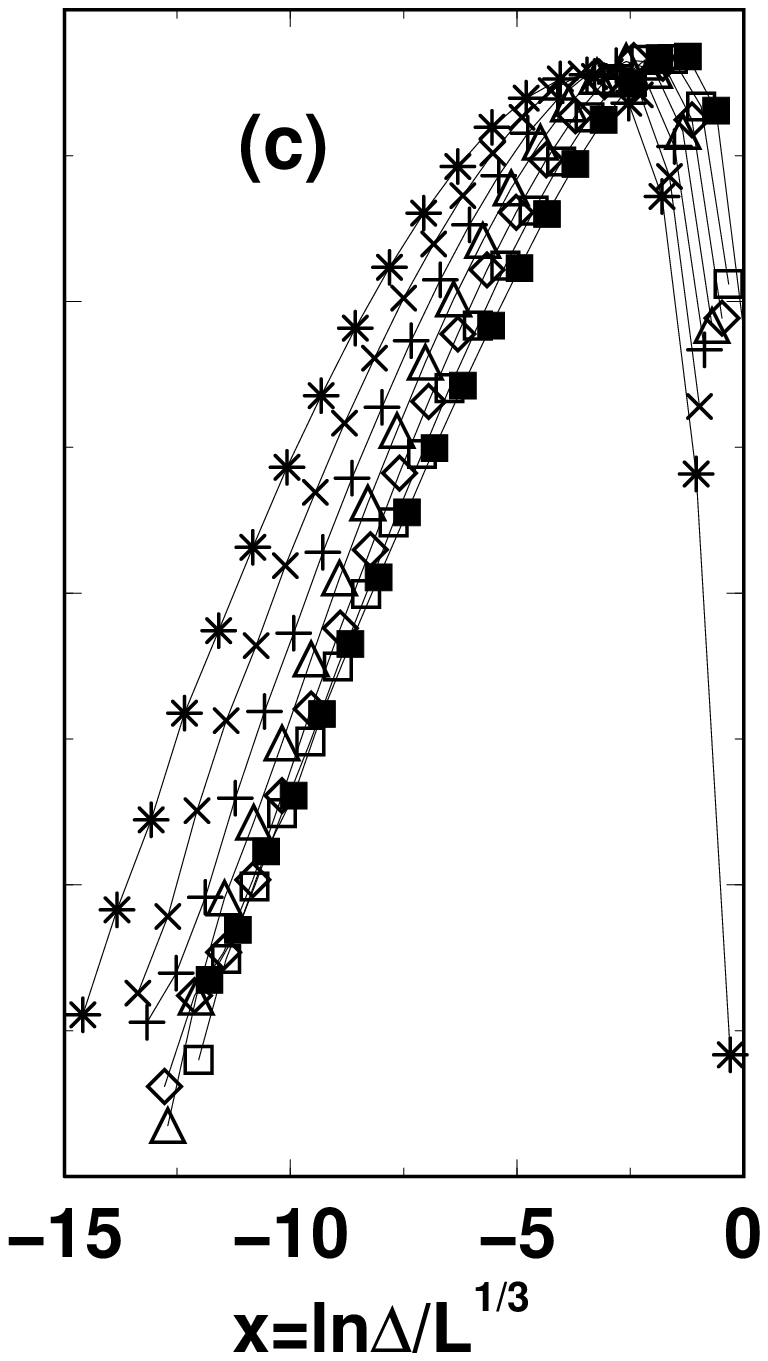}% Here is how to import EPS art

\caption{\label{fig:epsart11}  (a) Gap distribution generated by the original SDRG
  method at $\delta=1.0$. (b) The  scaling collapse assuming RSP
  and the scaling relation in  Eq.(\ref{distribut5}) with $\psi=1/2$, and 
   with $\psi=1/3$ (c). }

\end{figure}

The scaling collapse is tested also in the RSP.  The gap
distributions of the original SDRG are clearly broadening with increasing system sizes,
Fig. \ref{fig:epsart11}(a), the changes in the slopes
are more pronounced than in Fig. \ref{fig:epsart10},
indicating broadening without limit and the presence of RSP. The scaling
collapse  is not perfect for the
investigated sizes, although there is apparently a crossover phenomenon
present: for larger sizes the collapse is almost perfect, see
Fig.\ref{fig:epsart11}(b). 

In addition, the scaling collapse
is examined with and exponent $\psi=1/3$ as this value of exponent was
theoretically predicted in Ref. \refcite{Monthus,Monthusa} and numerically
found in Ref. \refcite{Lajko2} at the critical point, see Fig.\ref{fig:epsart11}(c). 
It can be argued that the scaling
collapse is far better with $\psi=1/2$ than with $\psi=1/3$. For $\psi=1/2$, the collapse is almost
perfect in the large-gap region and there is a systematic improvement for
larger sizes in the low-gap region. 
Very recently, $\psi=1/3$ was found with scaling investigation of
DMRG results\cite{Lajko2} at the same disorder for relatively small system sizes. The results presented here in Fig.\ref{fig:epsart11}
imply that the $1/3$ value of exponent is only an effective value due to a
crossover phenomenon. Since there is a uncertainty  regarding the value
of $\delta_c$, the scaling collapse of RSP was investigated at several
different strength of disorder in the $\delta \sim 0.7-1.5$ region. In this
region, the $1/2$ exponent gives always better scaling collapse than
the $1/3$ exponent.

We think that the presence of crossover phenomena, discussed  here in detail,
associated with changing finite-size corrections is the explanation why the
$S=1$ RAHC represents a very challenging problem often leading to
controversial results regarding the location of phase boundaries, see Section 7.5. in Ref. \refcite{Ferenc}
or Table II in Ref. \refcite{Lajko2}. This phenomenon is reported in detail in the 
$S=1/2$ RAHC in Ref. \refcite{Rieger,Riegera,Riegerb}, but also known in other
quantum spin chains.\cite{DMRG2a}

To summarize the scaling collapse investigation, scaling collapse
considerations should be carefully inspected as there are affected by
finite-size effect, while  the finite-size corrections allow a rather accurate estimates of different phases in terms of the slopes of the distributions in the low-gap regions.

\begin{figure}
\centering
\includegraphics*[width=6.65cm]{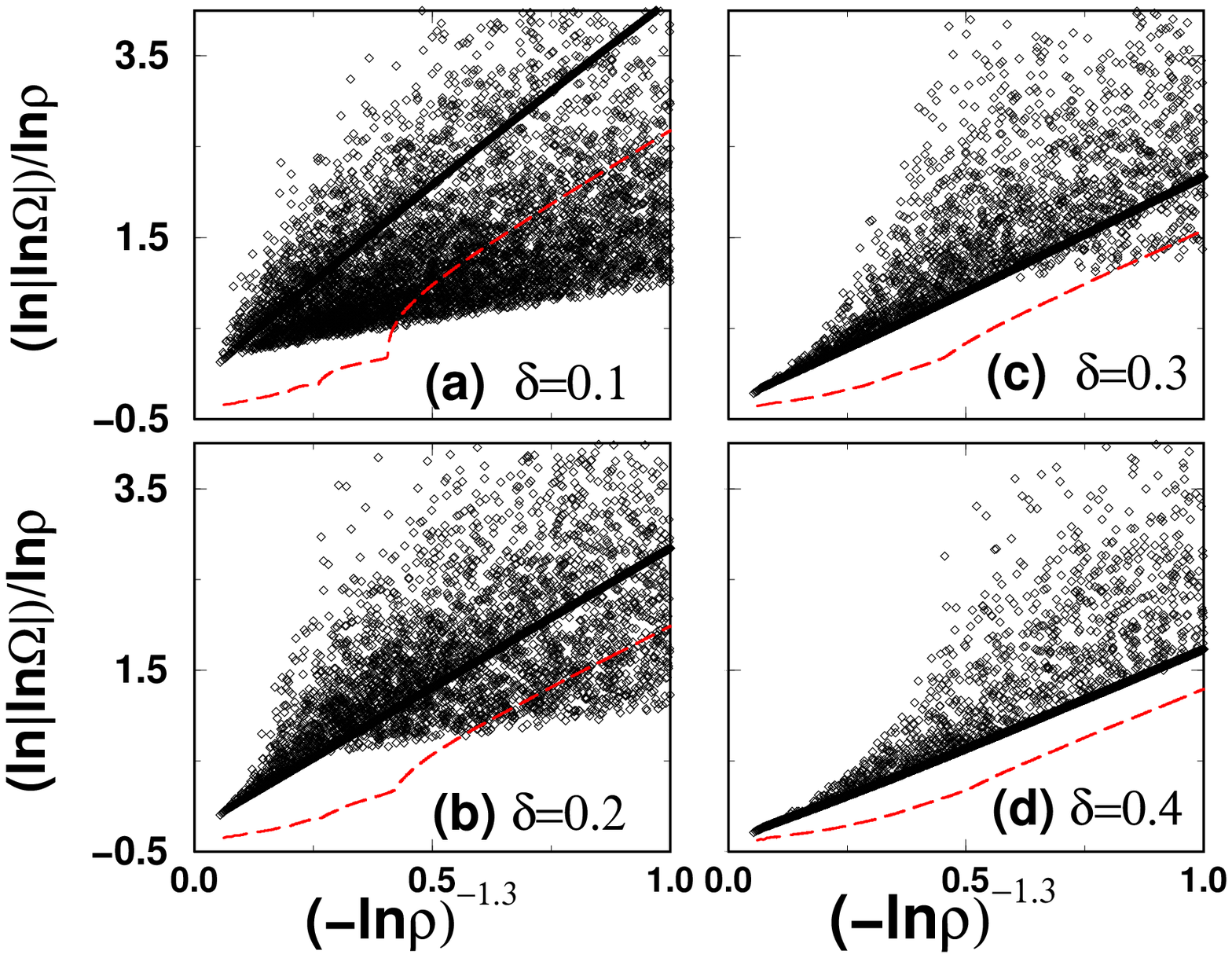}%
\hspace{0.1cm}%
\includegraphics*[width=5.95cm]{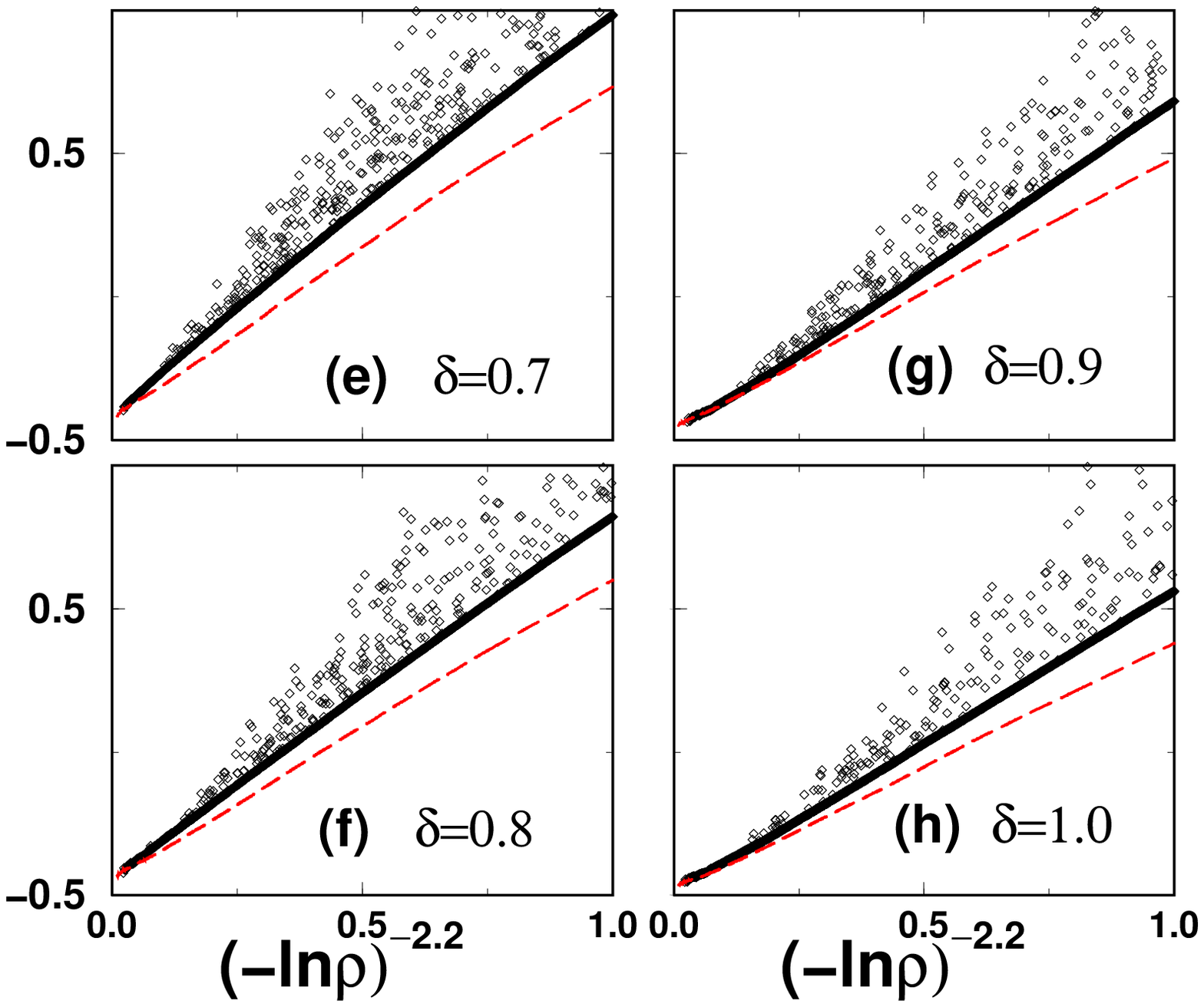}%
\caption{\label{fig:epsart13}  (Color online) Symbols denote the largest coupling
  applying the original SDRG method. Thick (red) staggered lines denote the same when
  the modified SDRG is used. Renormalization flows are depicted for
  $\delta=0.1, 0.2, 0.3, 0.4, 0.7, 0.8, 0.9, 1.0$ in Fig. (a)-(h),
  respectively. The $-1.3$ and $-2.2$ exponents in the label of the horizontal axis are
  determined in a fitting procedure.  }

\end{figure}

\subsection{Analysis of SDRG flows and the $\psi$ exponent}
SDRG and modified SDRG results were confronted also to relations in
                      Eq.(\ref{distribut3}) and (\ref{distribut2}) 
                      with the same conclusion as it is plotted in
Fig.\ref{fig:epsart2} at $\delta=1.0 $ disorder, but unfortunately also in a region below
this strength of disorder due to considerable errors. This indicates that probably  other
representations of RG flow are more proper to describe the phase diagram and the
appearance of the RSP. It is found that the best
quantity to be investigated is $\ln(|\ln\Omega|)/\ln\rho $ as a function of $\ln\rho$. From
Eq. (\ref{distribut3}), it is evident that assuming the presence of RSP this
                      product should tend to $-0.5$ in the
small-$\rho$ limit.

 The results of this investigation
are plotted in Fig.\ref{fig:epsart13} for both the original and modified SDRG cases. The data are
independent of the original length of the chain, which is 1 million in each
plotted case, they are practically free of finite-size corrections and thus allow a very precise estimate on $\psi$.  

For weak
disorder, very large fluctuations dominate the data, Fig.\ref{fig:epsart13},
which gradually decrease for stronger disorder. Around $\delta=0.2-0.4$ an
envelope curve appears below the fluctuating data points that 
allows a straight line fit for  small-$\rho$ estimate. 
 With
increasing disorder the small-$\rho$ estimates
gradually reach the $-0.5$ value, at $\delta=\delta_c$, and remains there for
stronger disorder. This representation provides a very accurate
prediction: RSP appears at $\delta_c=1.00(3)$ disorder. Notice that the small-$\rho$ estimates gradually decrease to
$-0.5$ and then with further increase of the strength of disorder they remain 
there, indicating clearly that the phase transition point is at
$\delta_c=1.00(3)$ where $\psi_c=0.50(1)$ but $\psi=0.5$ also deep in the RSP.

\section{Summary of the results and comparison to earlier findings}

In this paper, a non-perturbative GRG scheme and different variants of the perturbative SDRG
method  give a coherent picture about
the disorder-induced phase diagram of $S=1$ RAHC. 

The GRG method provides an
accountable phase diagram: for weak disorder, the system is in a Griffiths phase
with a disorder-dependent dynamical exponent; for stronger disorder the systems
reaches the RSP, at $\delta=\delta_c$,  where the dynamical exponent is infinity. The Griffiths phase can be
further divided, at $\delta=\delta_1$, into nonsingular and singular phases. GRG provides reasonable
estimates on the phase boundaries by applying finite-size scaling analysis,
see Fig.\ref{fig:epsart7}, $\delta_1=0.40(15)$ and $\delta_c=1.00(15)$. From
a theoretical point of view, the  GRG scheme is
well-established; from a numerical point of view, the data from GRG technique is well-justified.

 The original and modified SDRG  properly describe, in contrast to the earlier expectation, the
features of the system  and provide a high
accuracy for the phase boundaries in the large-$L$ limit:  $\delta_1=0.29(6)$ and $\delta_c=1.01(3)$,
in quantitative agreement with the latest works. The exponent $\psi$ is
shown out with different approaches and consistently found to be
$0.5$ in the RSP and in the critical point. 

The findings of this work are at least in qualitative agreement with all of
the earlier studies. This agreement can be considered, beyond the direct DMRG test at smaller sizes, as indirect test of the SDRG method in the asymptotic regime.

Good quantitative agreement can be observed with the latest numerical
investigations. The $\delta_1$ and $\delta_c$ phase boundaries, determined
in this work, are very close to the values in Ref. \refcite{Con,Todoa,Todob,Lajko2}. Hida's  early but hidden
estimate of $\delta_c$, see Fig.3 in Ref. \refcite{Hida}, is also close to
$1.01(3)$ that is the best estimate for $\delta_c$ in this work. The phase
boundary $\delta_1=0.29(6)$, deep in the Griffiths phase, well coincides
with independent estimates: $\delta_1=0.23$ (the relation of the
  power-law and box distribution is detailed in
  Ref. \refcite{Rieger}) in Ref. \refcite{Con} and $\delta_1=0.45$ in Ref. \refcite{Lajko2}.

There is a group of studies that successfully describe the phases in the $S=1$ RAHC by decomposing the $S=1$ spins into $S=1/2$ spins in an SDRG scheme. The success of these well-established methods is not questioned by the success of the original SDRG method; however,  with these earlier works only a
qualitative agreement is found, because these works contribute only qualitative
description of two phases RSP and a Griffiths phase\cite{S1} or they predict rather distinct phase boundaries: $\delta_c \sim 0.77$ is found in 
Ref. \refcite{Monthus,Monthusa} and
$\delta_c \sim 1.5$ in Ref. \refcite{Yanga}.

One of the earliest extended SDRG study\cite{Monthus,Monthusa} predicts distinct $\psi$
exponent in the critical point ($\psi_c=1/3$) and in the RSP ($\psi=0.5$). 
In the present work, a novel representation of the SDRG allows an accurate
numerical determination of $\delta_c$ and
$\psi$ exponent in the RSP as well as at the critical point, in both cases
$\psi=0.50(3)$ is observed, which is supported also by scaling collapse of the gap distributions
at and around $\delta_c=1.01(3)$. $\psi=0.5$ was found at $\delta=1.0$ also in a
series of recent QMC and SDRG studies.\cite{Con,Todoa,Todob,Berg} Thus, the properties of the RSP  at the critical point and for different 
strength of disorder seem to be coherent as it is
indicated in Ref. \refcite{Berg} by means of QMC.

The success of the SDRG approach, the fact that the perturbatively incorrect decimation
steps are not influencing the low-gap region of the gap
distribution, is an important contribution to our understanding of SDRG procedure. Furthermore, these accurate results are not
in contradiction with the conclusion
of Ref. \refcite{Boechat}, i.e., that the SDRG is not suitable to describe
 physical quantities like the free
energy. In this work, the study of the GRG and SDRG flows provide the physical
conclusion from the asymptotical behavior of the low-gap region within a usual RG framework without explicite
relevance on physical quantities like free energy.  Actually, this is obvious in SDRG context, see
Ref. \refcite{Ferenc}, \refcite{Vojta} or \refcite{Juhasz,Juhasza,Heiko,Monthus3,Monthus3a}.    

It is very likely that some of these new findings can be used to study the features of higher-$S$
RAHC and also other systems.\cite{Lajko} It would be instructive to test in what extent the phase diagram
for $S=1$ RAHC is reproducible by large scale QMC technique in
terms of physical quantities that are inaccessible by SDRG.  Moreover,
similar phases in the $S=1/2$ ferromagnetic-antiferromagnetic alternating
chain,\cite{Hida3,Hida3a} ladders,\cite{Arlego,Arlegoa} and whether the extended versions
of the SDRG, involving $S=1/2$ spins,  can be numerically  refined and can
provide quantitatively consistent results that remain subject of
further studies.

\section*{Acknowledgment}

The author is grateful to F. Igl\'oi for suggesting the 
meticulous investigation of SDRG method, to H. Rieger and
E. Carlon, G. F\'ath for cooperation on this field.  This work was supported by Kuwait University Research Grant No. SP. 09/02. 
 
%\section*{References} 

\end{document}